\documentclass[conference]{IEEEtran}
\IEEEoverridecommandlockouts
\usepackage[russian, main=english]{babel}
\usepackage{cite}
\usepackage{amsmath,amssymb,amsfonts}
\usepackage{algorithmic}
\usepackage{graphicx}
\usepackage{textcomp}
\usepackage{xcolor}
\usepackage{bigstrut}
\usepackage{caption}
\usepackage{subcaption}
\usepackage{flushend}

\def\BibTeX{{\rm B\kern-.05em{\sc i\kern-.025em b}\kern-.08em
    T\kern-.1667em\lower.7ex\hbox{E}\kern-.125emX}}
\begin{document}

\title{Comparison of Common Crawl News \& GDELT \\
{
\footnotesize 
}
}

\author{\IEEEauthorblockN{
Ameir El Ouadi and David Beskow}
\IEEEauthorblockA{\textit{Department of Systems Engineering} \\
\textit{United States Military Academy}\\
West Point, NY, USA \\
Email: ameir.elouadi@westpoint.edu}}

\maketitle

\begin{abstract}
The corpus of worldwide news is important for natural language processing, knowledge graphs, large language models, and other technical efforts.  Additionally, this corpus is important for understanding the people, places, organizations, and events that interact in real-time every day.  This paper compares two news datasets used for these tasks today, namely the Global Database of Events, Language, and Tone (GDELT) and Common Crawl News. Our research highlights the strengths and limitations of each dataset, analyzing their content and coverage. Notably, while GDELT relies on broadcasts, prints, and web news from across the globe, Common Crawl focuses on news sites from around the world gathered through web crawling. Our analysis revealed considerable differences in where the two datasets gather their news sources.
\end{abstract}


\section{Introduction}
The global stream of news data is a valuable source of information about the people, places, and events that interact daily worldwide.  Two prominent open-source datasets that claim to record and archive world news have emerged.  The first is the Global Database of Events, Language, and Tone (GDELT), and the other is the Common Crawl repository.  This paper will compare the coverage and content of Common Crawl and GDELT.  

The GDELT project is an open repository of news data, self-described as ``A Global Database of Society.'' The GDELT database collects broadcasts, prints, and web news from across the globe, spanning over 100 languages. GDELT’s data archive dates to January 1979.  GDELT's ability to monitor the world’s media provides context and information about international politics, which is crucial for analyzing society through a global lens. GDELT’s data provides a snapshot of global human society and what is happening in the news internationally on any given day. 

To handle the massive influx of information and data points coming from global events and news articles worldwide, GDELT relies on high-performance computing clusters and cloud storage solutions, specifically Google Cloud. The GDELT data files utilize Conflict and Mediation Event Observations (CAMEO) coding to classify events in the news articles automatically. The database identifies aspects of each news data observation, including people, locations, organizations, events, and images. The processed data is then stored in a structured format, allowing researchers to analyze events, themes, and emotions. GDELT's ability to capture and store large volumes of data facilitates comprehensive studies and insights into global events and trends.  It is important to note that the GDELT database provides metadata about news events and entities but not the raw text data itself.  To get the raw text data, a user must `re-hydrate' the URLs.  

The Common Crawl corpus is one of the most extensive data corpora available today, gathering and storing a snapshot of the accessible internet through web crawling. The Common Crawl Foundation collects and maintains an open repository of web crawl data containing raw web page data, metadata extracts, and text extracts. Common Crawl publishes this data as an open-source dataset for researchers, entrepreneurs, and developers.  Starting in 2016, the Common Crawl Foundation made a News stream available, which we will call CC-News. 

CCBot, a Nutch-based web crawler, has regularly crawled the internet and collected data for the Common Crawl dataset since 2007. The corpus contains over 250 billion pages of web data, adding up to five billion pages every month. The web crawl data is stored on Amazon Simple Storage Service (S3) as WARC files. Common Crawl data has proven valuable for research and training artificial intelligence, noting that 82\%  of raw tokens used to train GPT-3 come from the Common Crawl dataset. CC-News, a derived corpus of news data from Common Crawl, was used in this research. The CC-News dataset specifically contains articles from news sites all over the world.  This paper directly compares GDELT and CC-News, two publicly available and widely used news datasets. Analyzing each dataset, we seek to compare GDELT and CC-News while highlighting their strengths and limitations.

\section{Literature Review}
With the immense growth of the web over the past few decades, we have seen the increase in large-scale data and data repositories. The importance of large-scale data and its storage is especially relevant for research within computational journalism. The most significant advancements in computational journalism came in the 1960s with the invention of event data by Charles McClelland in his World Event/Interaction Survey (WEIS) \cite{mcclelland1976world}. Edward Azar’s Conflict and Peace Data Bank (COPDAB) followed WEIS. Both projects significantly contributed to international politics research efforts using news data \cite{azar1980conflict}.

In 1990, the National Institute of Standards and Technology (NIST) created the TREC program with funding from the U.S. Department of Defense Advanced Research Projects Agency (DARPA). The TREC program helped create test collections useful for research within computational journalism and organized conferences to disseminate the participants' research work. Early TREC campaigns heavily relied on news corpora for experimentation, especially utilizing full-text newswire articles \cite{sanderson2010test}. The primary focus of these tasks was ad-hoc retrieval, employing news corpora that usually consisted of hundreds of thousands of documents sourced from prominent news organizations \cite{sanderson2010test}.

Phillip A. Schrodt created the program and coding system, KEDS, responsible for the automated content analysis of news data and text to produce event data \cite{schrodt1994political}. In 2010, the U.S. Military sought to develop a crisis early warning system and funded the ICEWS project through DARPA, which utilized Schrodt’s KEDS as the coding basis \cite{o2010crisis}. ICEWS was a significant advancement in computational journalism and social science as it uses news data and text to be analyzed by systematical models for predicting global events, laying the groundwork for similar projects, such as GDELT. ICEWS also serves as a data repository of event data, one of the largest of its kind. 

Comparable to ICEWS, the GDELT project is also a data repository of event data, coded automatically from news articles worldwide \cite{leetaru2013gdelt}. GDELT differs from ICEWS in that it is open-source and available for anyone, as opposed to ICEWS, which can only be accessed for official government use. Since 2013, the GDELT dataset has been widely used for research within social science and was a groundbreaking achievement due to its accessibility. GDELT has been utilized in research in areas such as predicting social unrest \cite{qiao2017predicting}, analyzing global news coverage of disasters \cite{kwak2014first}, as well as analyzing fake news online \cite{vargo2018agenda}. Existing research also analyzes the GDELT dataset and compares it to other news and event repositories such as ICEWS \cite{ward2013comparing} and EventRegistry \cite{kwak2016two}.

The use of web corpora in computational journalism has been crucial for research. However, access to many datasets is restricted, which impedes many researchers. The accessibility and extensity of the Common Crawl dataset have allowed researchers to use it for many purposes, including natural language processing \cite{buck2014n}, machine learning \cite{smith2013dirt}, and creating derived data corpora \cite{mackenzie2020cc}. Previous research presents a preliminary analysis of the Common Crawl dataset regarding its scale, composition, and feasibility for research \cite{kolias2014exploratory}. In 2016, the Common Crawl project released the CC-News dataset, a crawl that specifically contains news data and publishes the data regularly shortly after the WARC files are written. Due to the nature of the web crawling process, Common Crawl datasets have been found to contain undesirable content, as presented in recent research \cite{luccioni2021s}. 


\section{Methodology and Results}

We collected both CC-News and GDELT datasets for the period from January 1, 2023, to June 31, 2023.  We then conducted temporal analysis, URL and domain name analysis, and language analysis of the two datasets.  One of the significant questions we wanted to analyze was to what extent the two datasets overlap.  If organizations must choose one of them, which contains the most content?  How do the two datasets differ in types of content?



\begin{figure}
\centering
\begin{minipage}{.5\columnwidth}
  \centering
  \includegraphics[width=0.96\linewidth,trim={2cm 2cm 1cm 2cm},clip]{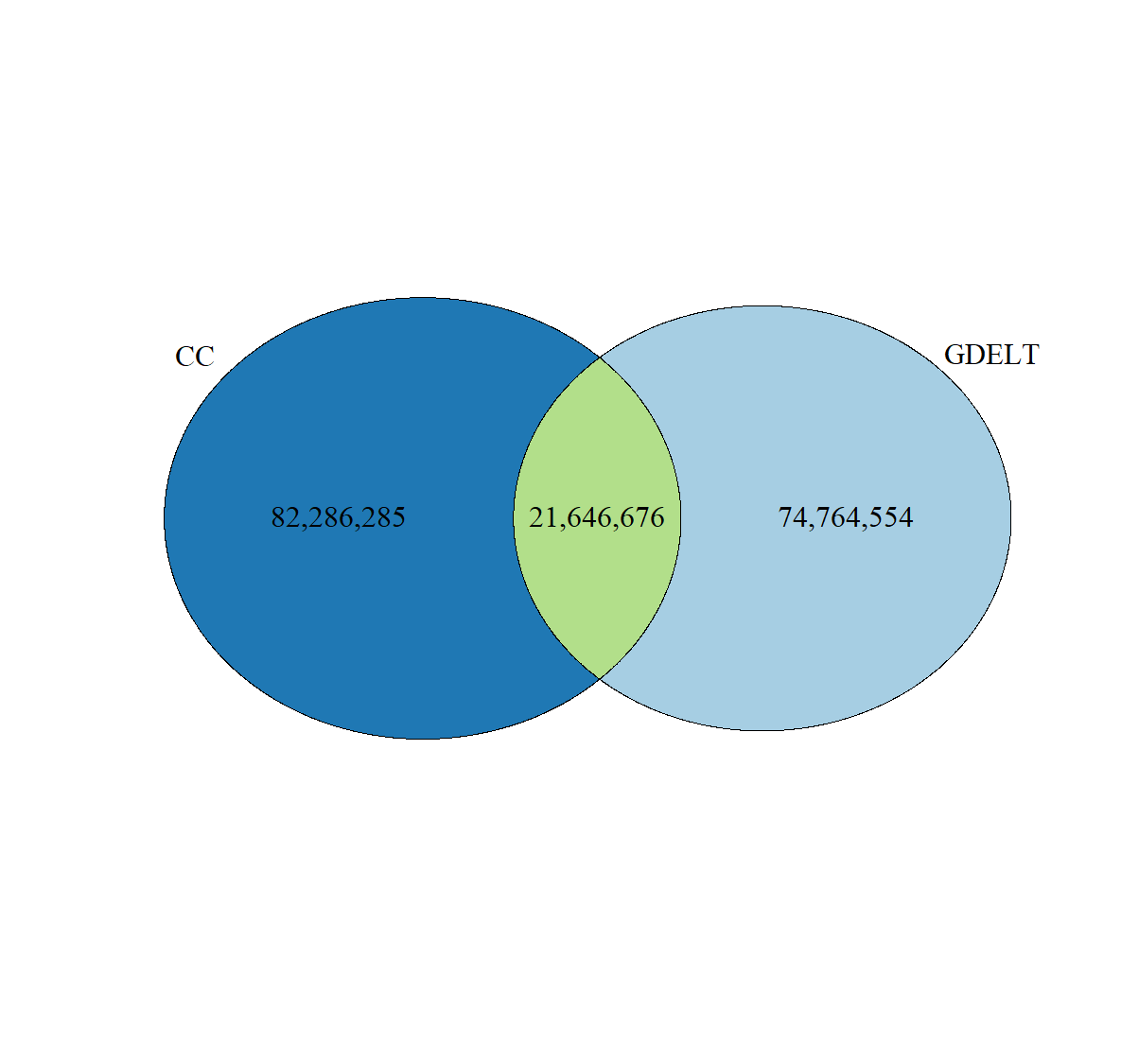}
  \captionof{figure}{Shared URLs}
  \label{fig:urlVenn}
\end{minipage}%
\begin{minipage}{.5\columnwidth}
  \centering
  \includegraphics[width=0.96\linewidth,trim={2cm 2cm 1cm 2cm},clip]{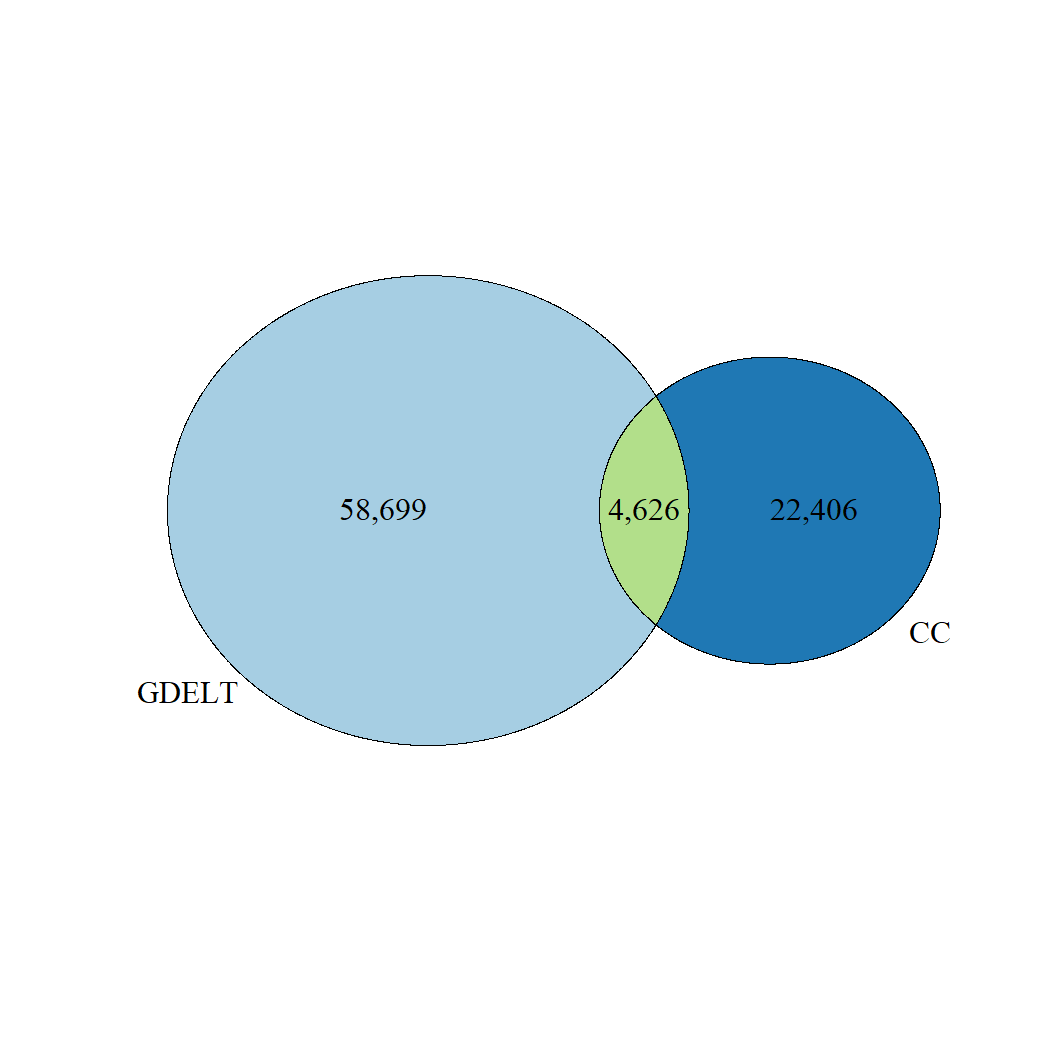}
  \captionof{figure}{Shared Domains}
  \label{fig:domainVenn}
\end{minipage}
\end{figure}

The scale of both datasets proved to be quite similar. Over the six months the data was collected, CC-News yielded 103,932,961 news articles, while GDELT yielded slightly less with 96,411,230 news articles. Figure \ref{fig:urlVenn} shows a Venn diagram showing the common news articles between the two datasets. GDELT data presents a more extensive range of news sources with 63,325 unique domains in the dataset. Despite its scale, CC-News contains less than half the unique domains of GDELT, with just 27,032. Comparing the domains of each dataset, we see that they only share 4,626 domains in common, as shown in Figure \ref{fig:domainVenn}, a Venn diagram of the domains of both datasets. This signifies that within GDELT and CC-News, only 5.4\% of news sources are shared. This is a noteworthy statistic, suggesting that the two datasets drastically differ in the news sources their articles are collected from.


We found only one domain in common when comparing each dataset's top ten most frequent domains (see Table \ref{tab:domains}). Within the top one hundred most frequent domains of each dataset, only six domains are shared between the two datasets: zazoom.it, itanews.yahoo.com, g1.globo.com, timesofindia.indiatimes.com, alarmeringen.nl, and udn.com. The disparity between the two datasets regarding shared news sources is substantial. Further demonstrating this phenomenon, only 17.11\% of CC-News domains are indexed in GDELT, and just 7.31\% of GDELT domains are indexed in CC-News. Despite being similar in size regarding the number of news articles, the two datasets showed stark differences in news sources.

The distribution of domains in each dataset differs slightly between GDELT and CC-News. In GDELT, the top ten most frequent domains account for 5.9\% of the entire dataset, with a large percentage of the dataset coming from domains contributing a small number of documents. Nearly half of the GDELT dataset, 46\%, comes from domains contributing ten or fewer news articles. The top ten most frequent domains in CC-News account for 7.7\% of the total corpus, similar to the percentage found in GDELT when analyzing the same statistic. Domains contributing ten or fewer articles in CC-News account for 28\% of the total corpus, a percentage considerably less than that of GDELT. 

In our analysis of the languages of the news articles in GDELT and CC-News, we found that a substantial percentage of articles from both datasets are written in English. The second most frequent language in the two datasets is Spanish. 36\% of articles in GDELT and 44\% of articles in CC-News are written in English. 
Both datasets share nearly the same group of languages, with eight of the top ten languages in common: English, Spanish, German, Russian, Italian, French, Arabic, and Turkish.

We found a considerable amount of non-news content after a deeper dive into the composition of domains within the CC-News dataset. For example, the domain www.nike.com contributes over 92,000 articles to the dataset. Most of these web pages from Nike, the shoewear company, can be loosely considered news as most are product announcements and press releases for their shoes and clothing lines. Reddit, the social news website and forum, contributed close to 20,000 articles to CC-News. Aside from non-traditional news content, we found that CC-News did contain some adult content-related domain names, such as www.xvideos.com and rt.pornhub.com. 

\begin{table}[htb]
  \centering
  \caption{Top Domains from GDELT (left) and CC-News (right)}
\resizebox{\columnwidth}{!}{\begin{tabular}{lcclcc}
	Top GDELT Domains&Lang&	Articles  & Top CC-News Domains& Lang & Articles \\
\hline 
	shuju.aweb.com.cn&zh&	996,647 &	entreprises.lefigaro.fr&fr&	1,144,893 \\
zazoom.it&it&		842,188 &	sportspyder.com&en&	1,071,322 \\
	businessdayonline.com&ja&	584,179 &	mirtesen.ru&ru&	1,007,584 \\
	msn.com&en&	562,012&	zazoom.it&it&	961,844 \\
	news.yahoo.com&en&	503,882 &	www.newsbreak.com&en&	860,118 \\
	inewsgr.com&el&	496,323 &	immobilier.lefigaro.fr&fr&	751,197 \\
	armenpress.am&hy&	479,918 &	familio.org&en&	610,376 \\
	chaspik.spb.ru&ru&	439,844 &	www.aktiencheck.de&de&	566,961 \\
	sohu.com&zh&	437,292 &	www.etcmagazine.com.ar&es&	539,394 \\
	haberler.com&tr&	369,094 &	tvsomalilandeurope.net&en&	443,993 \\
\hline
\end{tabular}}
\label{tab:domains}
\end{table}

\section{Conclusion and Future Work}
In an era where data plays a pivotal role in shaping decisions, solving problems, and advancing systems, the storage and accessibility of information are paramount. Web archiving fosters innovation and academic research in our digital age. However, many web archives are limited due to restricted public access or technological and copyright constraints. The GDELT project and Common Crawl have emerged as pioneers, offering extensive, freely available big data that contributes significantly to the academic community.

This paper presents a comparative analysis of two prominent news datasets: GDELT and CC-News. Our research examines various aspects of each dataset, including size, diversity, news source quality, and language distribution. Through this exploration, we provided beneficial insights to the computational journalism and social science communities. Researchers should weigh the distinct characteristics of these datasets when selecting datasets for their studies.

To conclude, GDELT, with its collection spanning traditional news sources worldwide, presents a comprehensive and diverse range of information within its news data. On the other hand, CC-News, derived from the Common Crawl, offers a higher volume of news data and provides a perspective on other web-based news content. We found the largest differences between the two datasets in their news sources. Furthermore, our comparative analysis highlights the strengths and limitations of each dataset, allowing researchers to make informed choices based on their specific study requirements, not limited to utilizing both datasets simultaneously to leverage the unique characteristics of each fully.

In future work, we will classify news articles by topic to better analyze the composition of the content of each dataset. It would also be helpful to understand how the emergence of large language models increases the value of these types of datasets and may limit their availability (for example, increasing numbers of news websites block CC-Bot).  As the value of these types of datasets increases, understanding the nuanced differences between them becomes important.  Our work takes a step toward exploring these differences. 

\vspace{0.2cm}\noindent\textbf{Author Note: }
The views expressed herein are those of the authors and do not reflect the position of the United States Military Academy, the Department of the Army, or the Department of Defense.\\

\bibliographystyle{ieeetr}
\bibliography{Paper}

@article{azar1980conflict,
  title={The conflict and peace data bank (COPDAB) project},
  author={Azar, Edward E},
  journal={Journal of Conflict Resolution},
  volume={24},
  number={1},
  pages={143--152},
  year={1980},
  publisher={Sage Publications Sage CA: Los Angeles, CA}
}

@inproceedings{buck2014n,
  title={N-gram Counts and Language Models from the Common Crawl.},
  author={Buck, Christian and Heafield, Kenneth and Van Ooyen, Bas},
  booktitle={LREC},
  volume={2},
  pages={4},
  year={2014}
}

@article{kolias2014exploratory,
  title={Exploratory analysis of a terabyte scale web corpus},
  author={Kolias, Vasilis and Anagnostopoulos, Ioannis and Kayafas, Eleftherios},
  journal={arXiv preprint arXiv:1409.5443},
  year={2014}
}

@inproceedings{kwak2014first,
  title={A first look at global news coverage of disasters by using the gdelt dataset},
  author={Kwak, Haewoon and An, Jisun},
  booktitle={Social Informatics: 6th International Conference, SocInfo 2014, Barcelona, Spain, November 11-13, 2014. Proceedings 6},
  pages={300--308},
  year={2014},
  organization={Springer}
}

@inproceedings{kwak2016two,
  title={Two tales of the world: Comparison of widely used world news datasets gdelt and eventregistry},
  author={Kwak, Haewoon and An, Jisun},
  booktitle={Proceedings of the International AAAI Conference on Web and Social Media},
  volume={10},
  number={1},
  pages={619--622},
  year={2016}
}

@inproceedings{leetaru2013gdelt,
  title={Gdelt: Global data on events, location, and tone, 1979--2012},
  author={Leetaru, Kalev and Schrodt, Philip A},
  booktitle={ISA annual convention},
  volume={2},
  number={4},
  pages={1--49},
  year={2013},
  organization={Citeseer}
}

@article{luccioni2021s,
  title={What's in the Box? A Preliminary Analysis of Undesirable Content in the Common Crawl Corpus},
  author={Luccioni, Alexandra Sasha and Viviano, Joseph D},
  journal={arXiv preprint arXiv:2105.02732},
  year={2021}
}

@article{o2010crisis,
  title={Crisis early warning and decision support: Contemporary approaches and thoughts on future research},
  author={O'brien, Sean P},
  journal={International studies review},
  volume={12},
  number={1},
  pages={87--104},
  year={2010},
  publisher={Blackwell Publishing Ltd Oxford, UK}
}

@inproceedings{mackenzie2020cc,
  title={CC-News-En: A large English news corpus},
  author={Mackenzie, Joel and Benham, Rodger and Petri, Matthias and Trippas, Johanne R and Culpepper, J Shane and Moffat, Alistair},
  booktitle={Proceedings of the 29th ACM International Conference on Information \& Knowledge Management},
  pages={3077--3084},
  year={2020}
}

@misc{mcclelland1976world,
  title={World event/interaction survey codebook},
  author={McClelland, Charles A},
  year={1976},
  publisher={ICPSR Ann Arbor}
}

@article{qiao2017predicting,
  title={Predicting social unrest events with hidden Markov models using GDELT},
  author={Qiao, Fengcai and Li, Pei and Zhang, Xin and Ding, Zhaoyun and Cheng, Jiajun and Wang, Hui and others},
  journal={Discrete Dynamics in Nature and Society},
  volume={2017},
  year={2017},
  publisher={Hindawi}
}

@article{sanderson2010test,
  title={Test collection based evaluation of information retrieval systems},
  author={Sanderson, Mark and others},
  journal={Foundations and Trends{\textregistered} in Information Retrieval},
  volume={4},
  number={4},
  pages={247--375},
  year={2010},
  publisher={Now Publishers, Inc.}
}

@article{schrodt1994political,
  title={Political science: KEDS—a program for the machine coding of event data},
  author={Schrodt, Philip A and Davis, Shannon G and Weddle, Judith L},
  journal={Social Science Computer Review},
  volume={12},
  number={4},
  pages={561--587},
  year={1994},
  publisher={Sage Publications Sage CA: Thousand Oaks, CA}
}

@inproceedings{smith2013dirt,
  title={Dirt cheap web-scale parallel text from the common crawl},
  author={Smith, Jason R and Saint-Amand, Herve and Plamada, Magdalena and Koehn, Philipp and Callison-Burch, Chris and Lopez, Adam},
  year={2013},
  organization={Association for Computational Linguistics}
}

@article{vargo2018agenda,
  title={The agenda-setting power of fake news: A big data analysis of the online media landscape from 2014 to 2016},
  author={Vargo, Chris J and Guo, Lei and Amazeen, Michelle A},
  journal={New media \& society},
  volume={20},
  number={5},
  pages={2028--2049},
  year={2018},
  publisher={Sage Publications Sage UK: London, England}
}

@article{ward2013comparing,
  title={Comparing GDELT and ICEWS event data},
  author={Ward, Michael D and Beger, Andreas and Cutler, Josh and Dickenson, Matthew and Dorff, Cassy and Radford, Ben},
  journal={Analysis},
  volume={21},
  number={1},
  pages={267--297},
  year={2013}
}

\end{document}